\documentclass[onecolumn,preprintnumbers,12pt,amsmath,amsssymb]{revtex4}
\usepackage{graphicx}
\def\gsim{ \lower .75ex \hbox{$\sim$} \llap{\raise .27ex \hbox{$>$}} }
\def\lsim{ \lower .75ex \hbox{$\sim$} \llap{\raise .27ex \hbox{$<$}} }
\def\be{\begin{equation}}
\def\ee{\end{equation}}
\def\bea{\begin{eqnarray}}
\def\eea{\end{eqnarray}}
\def\lf{\left (}
\def\rt{\right )}

\begin{document}
\preprint{CU-TP-1167}
\preprint{IUCAA-55/2006}
\title{Mach's Holographic Principle}
\author{Justin Khoury$^{1}$ and Maulik Parikh$^{2,3}$}

\affiliation{$^1$Perimeter Institute for Theoretical Physics, 
31 Caroline St. N., Waterloo, ON, N2L 2Y5, Canada \\ 
$^{2}$Institute for Strings, Cosmology and Astroparticle Physics,
  Columbia University, New York, NY 10027, USA \\
$^{3}$Inter-University Centre for Astronomy and Astrophysics, 
Post Bag 4, Pune 411007, India} 

\begin{abstract}
\begin{center}
{\bf Abstract}
\end{center}
\noindent
Mach's principle is the concept that inertial frames are
determined by matter. We propose and implement a
precise formulation of Mach's principle in which matter and geometry
are in one-to-one correspondence. Einstein's equations are not
modified and no selection principle is applied to their
solutions; Mach's principle is realized wholly within 
Einstein's general theory of relativity. 
The key insight is the observation that, in addition to bulk matter,
one can also add boundary matter. Specification of both
boundary and bulk stress tensors uniquely specifies the geometry and 
thereby the inertial frames. Our framework is similar to that of the
black hole membrane paradigm and, in asymptotically AdS space-times, 
is consistent with holographic duality. 
\end{abstract}

\maketitle

\section{Mach's Principle}

Acceleration appears absolute. A snapshot of a rotating bucket of water
reveals, through the gentle curve in the water's surface, that the
bucket was rotating. Two rocks tied with a rope and set spinning about
an axis perpendicular to the rope are measurably distinct from the
same two rocks undergoing linear motion: the rope becomes tense.
A passenger in an elevator or a windowless spaceship is aware of
starts and stops even though the vehicle is a closed system.

With his principle of equivalence, Einstein recognized that gravity was simply 
acceleration in disguise. Moreover, Einstein's equations, like Newton's law of
gravitation, indicate that matter is the source for gravity. But if
acceleration and gravity are linked, and if gravity depends on
matter, then can acceleration be attributed to matter?

This imprecise notion is the essence
of Mach's principle, which asserts that whether the motion of a given observer is inertial or not is determined by ``the distant stars," Mach's memorable phrase for the matter distribution in the universe~\cite{mach}.
If this profound claim were true, all motion, not just inertial motion, would be
relative. Acceleration would not be absolute, for to accelerate
without matter would be meaningless: there would be nothing to
accelerate with respect to. The water in a bucket ``rotating" all alone in the universe would not rise up at the sides, as there would be no sense in which a solitary bucket could be said to be rotating. As Weinberg poetically points out~\cite{weinberg}, to appreciate the significance of Mach's principle one need only perform a pirouette underneath a starry sky. Is it mere coincidence that the frame in which one's arms fly outwards is the same as the frame in which the distant stars appear to spin overhead, or is there a deeper dynamical explanation?

In his landmark paper on the foundations of general relativity \cite{foundation}, Einstein sought to make the relativity of all motion one of the cornerstones of his new theory. 
But ironically, general relativity did not in the end seem to support Mach's idea. In general relativity whether a given world-line is inertial or accelerating depends on whether or not it satisfies the geodesic equation. This in turn depends on the metric which, indeed, is related through Einstein's equations to the matter distribution, encoded in the stress-energy tensor. However, the point is that ultimately the metric exists whether or not there is matter present. The existence of Minkowski space most emphatically underlines this point: geodesics and inertial frames exist even in the total absence of all matter. Although there are several distinct versions of what is meant by Mach's principle, the example of Minkowski space establishes that one common interpretation --- that inertial frames here and now are determined by some kind of averaging over matter elsewhere --- can immediately be ruled out.

In 1918 Einstein proposed a different definition of Mach's principle:
\begin{quote}
Mach's Principle: The $G$-field is without remainder determined by the
masses of bodies. Since mass and energy are, according to results of
the special theory of relativity, the same, and since energy is
formally described by the symmetric energy tensor ($T_{\mu \nu}$),
this therefore entails that the $G$-field be conditioned and determined
by the energy tensor~\cite{einstein}.
\end{quote}

\noindent That is, Mach's principle holds if the metric (the ``$G$-field'') is,
up to diffeomorphisms, uniquely specified by the stress tensor. In Einstein's 1918 formulation, there is a one-to-one correspondence between matter and
geometry. This formulation would accept that in Minkowski space inertial frames exist without matter, but it would require that Minkowski space be the unique empty space-time. If Minkowski space were the unique space-time devoid of matter, one might still be able to claim that specifying the matter distribution somehow specifies the inertial frames. But of course there exist besides Minkowski space a host of perfectly fine solutions to the vacuum Einstein equations, among which are several well-known exact solutions like the Schwarzschild and Kerr black holes. Evidently, Einstein's 1918 version of Mach's principle is also in trouble. 

Indeed, that matter and geometry are, contrary to Einstein's 1918 proposal,
not entirely in one-to-one correspondence can be seen in a variety of
ways: 

\begin{itemize}

\item The Weyl tensor is not determined by matter.

Einstein's equations determine the Ricci tensor in terms of the
stress tensor:
\begin{equation}
R_{\alpha\beta} = 8 \pi G_D\lf T_{\alpha\beta} - \frac{1}{D-2}T g_{\alpha\beta} \rt \; .
\end{equation}
But the complete geometry is encoded in the Riemann tensor which, in four or more dimensions, 
includes not only the Ricci tensor but also
the Weyl tensor. And, unlike the Ricci tensor, the Weyl tensor is independent of matter (give or take a Bianchi identity). Consequently, matter --- $T_{\alpha\beta}$ --- does not fully determine the Riemann tensor, and hence the geometry.

\item General relativity permits gravitational waves.

A more physical way of stating the problem is to note that general
relativity permits gravitational waves. But these can exist as
independent fluctuations even in empty space-time; there exist
gauge-invariant solutions to the homogeneous wave equation.

\item Einstein's equations need boundary conditions.

Since Einstein's equations are second-order partial differential
equations, to obtain a unique solution one needs to supplement them by boundary/initial conditions for the metric. These are usually in the form of an
induced metric $h_{\alpha\beta}$ and extrinsic curvature $K_{\alpha\beta}$ for
some appropriate hypersurface. The boundary conditions are arbitrary and are also apparently independent of matter.

\end{itemize}

\noindent For all these reasons, Einstein's 1918 version of Mach's principle does not 
seem to hold. A related embarrassment from a Machian perspective is the
existence of solutions with nonzero global angular momentum; Mach's
ideas on the relativity of all motion imply that a closed system
cannot be rotating --- rotating with respect to what? --- but in the
Kerr black hole, as well as in other examples, general relativity
permits solutions that have non-vanishing total angular momentum. 

In attempts~\cite{barbour,wheeler} to save Mach's principle, two separate lines of attack have been pursued. According to one, favored initially
by Einstein himself, general relativity is preserved intact but a selection
rule is imposed on the space of solutions. Einstein, for example, demanded
that cosmologies have compact spatial topology. There are several
drawbacks to this approach, not the least its {\it ad hoc} nature. For instance, demanding compact spatial topology rules out Minkowski space, while $R\times T^3$ with arbitrarily large torus is allowed. Furthermore, it fails to eliminate the problem of boundary conditions or of gravitational waves.
The second line of attack consists of modifying general
relativity. This approach has also not worked. Indeed, so long as the dynamics of gravity are governed by a differential equation, arbitrary and apparently matter-independent boundary conditions are needed. Nor have other variants of Mach's principle met with great success. Thus it would seem that Mach's principle is one of those tantalizingly beautiful ideas that sadly are not realized in nature.

Nevertheless, it is the purpose of this paper to argue that Mach's ideas can be
precisely and concretely implemented within Einstein's general theory
of relativity. The version of Mach's principle we will implement is Einstein's 1918 formulation. General relativity, we shall see, can be rewritten so as to make the geometry depend entirely
and uniquely on the stress tensor. The equations are not modified in any way. Instead, we shall show that there exists a recasting of the theory that is entirely consistent with Mach's principle.

To understand how this is possible, let us note that all
the aforementioned objections --- the matter-independence of the Weyl
tensor, the existence of gravitational waves, solutions with global
rotation --- can be related to one thing: the need for boundary
conditions. To reiterate, Einstein's equations are second-order
differential equations for the metric; to determine their solution
they need to be supplemented by boundary conditions. This, then, is
the crux of the problem. Boundary conditions are needed. The stress
tensor is not enough.

Yet this way of stating the problem also points to an unexplored
loop-hole. In recent years, there has been renewed interest in
boundary matter. Boundary matter had largely been neglected in earlier approaches to Mach's principle, perhaps because it seemed too exotic at the time. But boundary matter is part of the bread and butter of theoretical physics today, and arises in a variety of 
contexts. It appears in brane-world scenarios, in which our universe is
embedded in a higher-dimensional space. It appears in Ho\v rava-Witten
constructions as end-of-the-world branes~\cite{horavawitten}. It appears again in AdS/CFT
as holographically dual matter~\cite{magoo}. It appears in the Brown-York construction of a boundary stress tensor
for the gravitational field~\cite{brownyork}. And, perhaps most relevantly, boundary
matter appears in the black hole membrane paradigm~\cite{membraneparadigm}, where the matter lives on the black hole horizon, an internal boundary of space-time for an external observer. 

Adding a stress tensor at a boundary does not affect the bulk Einstein
equations. Moreover, even space-times that are empty in the bulk, such
as Minkowski space, may admit stress tensors living on their
boundaries. Can the bulk metric --- and thus the set of geodesics --- be
extracted by specification of {\it both} bulk and boundary stress tensors?
This paper will answer that question in the affirmative. We will see
that, by allowing for two separate boundary stress tensors, every
gauge-equivalent class of metrics can be mapped to a matter
distribution. In the process, Einstein's equations are
retained and no solutions are sacrificed.

Trading boundary conditions for appropriate sources is an oft-employed technique in physics.
An example that immediately springs to mind is the method of image charges in electrostatics in which conducting
boundary conditions are replaced by fictitious charges. A closer analogue to our proposal is the membrane description of black hole
horizons~\cite{membraneparadigm,membraneaction}. The membrane paradigm is the remarkable notion that, from the perspective of
an outside observer, a black hole behaves precisely as if it were cloaked in a fluid membrane living at the event horizon. That is, the equations of motion of fields in the background of a black hole, with regular boundary conditions at the horizon, can be rewritten so that the same equations describe the fields interacting with a source at the horizon --- a membrane. Yet, the membrane approach is more than a mathematical trick. To an observer hovering outside the horizon, the membrane appears to behave like a real, dynamical fluid. It conducts electricity according to Ohm's law; it generates heat through Joule's law; and it flows following the Navier-Stokes equation. Only by jumping into the black hole can the observer realize the illusory nature of the membrane.

In some sense our proposal can be viewed as an ``inside-out" version of the membrane paradigm, a parallel most apt for space-times 
with causal horizons. Take de Sitter space. The natural location for the boundary stress energy in this case is
on the ``stretched" horizon, a time-like surface hovering just inside the causal horizon. Much like the black hole case, this membrane is dynamical and satisfies a host of
classical equations. Intriguingly its surface energy density inferred from Israel-like junction conditions has the equation of state of dust. To a bulk observer
this boundary dust plays the role of the distant stars, relative to which accelerated motion in the bulk can meaningfully be defined.

Moreover, our boundary matter has a compelling interpretation in terms of the Brown-York stress energy~\cite{brownyork} of the gravitational field in the interior and exterior regions of space-time. It had long been conjectured that general relativity could be proven to be Machian by somehow taking into account gravitational stress energy. But of course a local notion of stress energy for gravity is meaningless. Stress tensors are usually constructed from fields and first derivatives of fields, both of which can be made trivial at any point by a suitable choice of coordinates. Instead, as argued by Brown and York using Hamilton-Jacobi theory, the natural location for gravitational stress energy is at the boundary. 

Finally, in asymptotically AdS space-times our proposal is consistent with the holographic correspondence. (Intriguingly, a connection between Mach's principle and holography has also been made in Ho\v rava's Chern-Simons M-theory \cite{horava}.) The Brown-York stress tensors mentioned above are now understood as the holographic stress tensors for the dual field theory in one lower dimension. Indeed, the correspondence identifies the radial direction (say in ADM coordinates) 
in an asymptotically AdS space-time with the RG scale of the dual theory~\cite{holoRG1, holoRG2, holoRG3}.
Placing an effective boundary at some radial location therefore represents
an ultraviolet cut-off in the dual theory where RG initial conditions can be specified. Furthermore, replacing the exterior region with matter on the boundary
corresponds in the dual language to integrating out high-energy degrees of freedom, whose quantum stress-energy tensor is just the Brown-York tensor for the exterior gravitational field.
Similarly, the Brown-York tensor for the interior is the stress tensor for the low-energy degrees of freedom. This interpretation also makes it clear that there is nothing special
about the location of the boundary. In the Wilsonian sense, shifting the boundary along the radial direction simply corresponds to choosing a different RG cut-off. 

To summarize, we implement Mach's principle within general relativity in a way that ties in with the membrane paradigm, the Brown-York notion of stress energy
for the gravitational field, and the holographic correspondence. Our proposal amounts to a rewriting of Einstein's theory, combined with a rule for obtaining the boundary stress tensors, that together make Mach's principle manifest. This reformulation does not affect Einstein's equations; indeed, any formulation of Mach's principle that was {\it not} consistent with those equations would already be in trouble. 

Often in physics the recasting of an existing theory has deepened our understanding of nature.
Consider again the black hole membrane paradigm. While superficially just a rewriting of classical equations, the key insight that the event horizon behaves as a dynamical fluid led to a host of conceptual breakthroughs: black hole entropy was understood as a local property; astrophysical phenomena like the Blandford-Znajek process~\cite{blandford} were clarified;
the no-hair theorem became intuitive; and the complementarity principle was motivated as an approach to the information
puzzle~\cite{susskind,thooft}. The rewritten equations are deservedly termed a new paradigm. We hope that our framework may similarly help to shed new light on old problems, such as the origin of noninertial forces and the relativity of all motion.

A brief outline of this work is as follows. In Section~\ref{bc} we argue that the
various obstacles in the way of a realization of Mach's principle all
boil down to the need for boundary conditions. Section~\ref{bdy} presents an electromagnetic counterpart to Mach's principle; the boundary conditions are encoded in charges and currents living on a kind of Faraday
cage. We then show that the boundary conditions for gravity too can be regarded as originating in boundary sources, and we propose a particular kind of boundary stress tensor that encodes the
boundary conditions. In Section~\ref{charge} we show that our particular prescription for obtaining the boundary stress tensor also takes care of the problem of net global angular momentum by precisely canceling any global rotation of the space-time. In Section~\ref{examples} we apply our proposal for boundary stress tensors to a variety of well-known space-times and we read off the form of the boundary matter. Remarkably the boundary matter typically turns out to be simply pressureless dust. Section~\ref{comp} contrasts our proposal with earlier attempts to reconcile Mach's principle with gravity. We conclude in Section~\ref{summary} with a brief summary and some directions for future work.

\section{Boundary Conditions} \label{bc}

Three reasons for the apparent failure of Mach's principle is that general relativity admits
gravity waves, the geometry is encoded partly in the Weyl tensor,
and Einstein's equations are subject to boundary conditions. In
this section we argue that these three objections are in fact equivalent. This
will motivate us to cast the problem entirely in terms of boundary
conditions, in preparation for the next section wherein we will capture
those boundary conditions through the addition of boundary matter.

We begin by reviewing the relation between gravitational wave
solutions and the Weyl tensor in $D>3$ space-time dimensions. 
First note that the space-time geometry
within any coordinate patch is determined either by specifying the local metric
or by specifying the local Riemann curvature. Of course, the Riemann
tensor is easily computed from the metric. To see that the converse
is also true, note that one can express the metric locally in terms of the Riemann tensor as
\be
g_{\alpha\beta}(x) = \eta_{\alpha\beta} - \frac{1}{3}
R_{\alpha\gamma\beta\delta} x^\gamma x^\delta + \ldots\,,
\ee
where the $x$'s are by definition Riemann normal coordinates. Thus for our purposes it suffices to focus on the Riemann tensor.

Now the part of the Riemann tensor that does not figure in the Einstein equations
is the Weyl tensor, which, again in $D>3$ space-time dimensions,
is related to the Riemann and Ricci tensors via
\be
C_{\alpha\beta\gamma\delta} = R_{\alpha\beta\gamma\delta} + \frac{2}{D-2} \lf g_{\alpha[\delta}R_{\gamma]\beta} +
g_{\beta[\gamma}R_{\delta]\alpha} \rt + \frac{2}{(D-1) (D-2)} Rg_{\alpha[\gamma}g_{\delta]\beta} \,,
\ee
where the commutator is normalized according to $x_{[a,b]} \equiv (x_{ab} - x_{ba})/2$. Since Riemann is uniquely mapped to the gauge-equivalent class of the metric, and since Weyl is unspecified by the local stress tensor, it follows that homogeneous solutions to Einstein's equations --- gravity waves --- are encoded in Weyl. Indeed, one can make this more explicit by noting that the Bianchi identity, $\nabla_{[\alpha}R_{\beta\gamma]\delta}^{\;\;\;\;\;\;\;\epsilon} = 0$, combined with Einstein's equations, implies a constraint on Weyl:
\be
\nabla^\delta C_{\alpha\beta\gamma\delta} =  16 \pi G \lf \frac{D-3}{D-2} \rt \lf \nabla_{[\beta}T_{\alpha]\gamma} - \frac{1}{D-1}
g_{\gamma[\alpha}  \nabla_{\beta]}T \rt \; .
\label{weyldiv}
\ee
Then, after some work ~\cite{penroserindler}, it is possible to show that 
the linearized version of this equation can be repackaged as the wave equation for a massless spin-2 field, establishing the anticipated relation between the Weyl tensor and gravitational waves. Incidentally,~(\ref{weyldiv}) implies that the Weyl tensor is not entirely independent of bulk matter. Since the stress tensor only appears acted upon with derivatives, however, qualitatively the Weyl tensor at a point encodes the part of the curvature due to matter elsewhere~\cite{hawkingellis}. 

Having motivated the equivalence between gravity waves and Weyl, we turn
to Weyl versus boundary conditions. Solving the Einstein equations requires boundary
conditions for the metric, for instance the induced metric
$h_{\alpha\beta}$ and extrinsic curvature $K_{\alpha\beta}$ of a co-dimension one
hypersurface. For a time-like boundary with unit normal $n^\alpha$ (with mostly positive metric signature), these are given by
\bea
\nonumber
h_{\alpha\beta} & = & g_{\alpha\beta} - n_\alpha n_\beta\;; \\
K_{\alpha\beta} & = & h_\alpha^{\;\;\gamma}\nabla_\gamma n_\beta\,.
\eea
Meanwhile,~(\ref{weyldiv}) is a
first-order differential equation for Weyl. This means that, given some bulk
matter, a unique solution is obtained by specifying one boundary
condition: the boundary value of the Weyl tensor. Thus it remains to
establish a one-to-one correspondence between the boundary value of the Weyl tensor and the choice of $h_{\alpha\beta}$ and $K_{\alpha\beta}$.

To do so, we assume for simplicity an empty bulk,
$T_{\alpha\beta}=0$. The bulk Weyl tensor evaluated at the boundary
can be decomposed into its electric and magnetic parts:
\bea
\nonumber
E_{\mu\nu} & \equiv & C_{\alpha\beta\gamma\delta}n^\alpha n^\gamma h_\mu^{\;\beta}h_\nu^{\;\delta}\;; \\
B_{\mu\nu\gamma} & \equiv & C_{\alpha\beta\gamma\delta}h_\mu^{\;\alpha}h_\nu^{\;\beta}n^\delta\,.
\eea
Both are traceless, $E^\mu_{\;\mu}=0$ and $B^\alpha_{\;\nu\alpha}=0$, and have the following symmetries: $E_{\mu\nu} = E_{\nu\mu}$; $B_{\alpha\beta\gamma}=-B_{\beta\alpha\gamma}$; $B_{[\alpha\beta\gamma]}=0$.

To derive a relation between $E$ and $(h,K)$, we take the trace, over $\alpha$ and $\gamma$ of the Gauss relation, 
$^{(D-1)}R_{\alpha\beta\gamma\delta} =
h_\alpha^{\;\kappa}h_\beta^{\;\mu}h_\gamma^{\;\nu}h_\delta^{\;\sigma}
R_{\kappa\mu\nu\sigma}+K_{\alpha\gamma}K_{\beta\delta}-
K_{\beta\gamma}K_{\alpha\delta}$, and use $R_{\alpha\beta}=0$ to obtain
\be
E_{\mu\nu} = -^{(D-1)}R_{\mu\nu} +
KK_{\mu\nu}-K_\mu^{\;\alpha}K_{\nu\alpha}\,.
\label{Eexp}
\ee
Next, take the identity $R_{\alpha\beta\gamma\delta}n^\delta =
(\nabla_\alpha\nabla_\beta-\nabla_\beta\nabla_\alpha)n_\gamma$. Contracting with $h_\mu^{\;\alpha}h_\nu^{\;\beta}$
and substituting the expression for the Weyl tensor and the extrinsic curvature, we find
\be
B_{\mu\nu\gamma} = D_\mu K_{\nu\gamma} - D_\nu K_{\mu\gamma} \,,
\label{Bexp}
\ee
where $D_\mu$ is the covariant derivative associated with $h$. Since $E$ and $B$ are traceless, taking the trace of the above expressions incidentally gives the usual initial-value constraints of general relativity.
Equations~(\ref{Eexp}) and~(\ref{Bexp}) display the explicit map
between the boundary data $h$ and $K$, and the boundary value of the Weyl tensor. By ~(\ref{weyldiv}), the components of the Weyl tensor at some point in the bulk are then also implicitly functions of the boundary data, which is what we wanted to show. 

As a check on the number of degrees of freedom, note that $h$ and $K$
are symmetric tensors in $D-1$ dimensions, for a total of $D(D-1)$
components; each is covariantly conserved:
$D^\mu h_{\mu\nu} = 0$ and $D^\mu K_{\mu\nu} - D_\nu K =0$, bringing
the total down to $D(D-3)+2$; the trace of~(\ref{Eexp}) gives the
Hamiltonian constraint $^{(D-1)}R =
K^2-K_{\alpha\beta}K^{\alpha\beta}$; and the final condition
corresponds roughly speaking to the usual freedom in specifying the
co-dimension one hypersurface, which requires one constraint. 
This leaves us with $D(D-3)$
arbitrary gauge-invariant degrees of freedom at the boundary which, to complete the circle, is
indeed twice the number of graviton polarizations in $D$ dimensions.

\section{The Distant Stars: Matter at the Boundary of Space} \label{bdy}

In this section we show that boundary conditions for the metric can
be uniquely specified in terms of boundary stress tensors. This
fulfills Einstein's statement of Mach's principle --- that
the metric field be uniquely determined by matter stress energy --- as
long as ``matter'' refers to both bulk and boundary stress tensors. To illustrate our procedure we first consider a related problem in electrodynamics.

\subsection{Electromagnetic analogy} \label{EM}

There exists an electromagnetic counterpart to the version of Mach's principle
presented here. Asking whether the metric --- the gravitational field
--- is determined entirely by matter sources is akin to asking
whether the electromagnetic field is entirely determined by
electromagnetic sources --- charges and currents. The answer is no
because, once again, there are boundary conditions. Alternatively, we note 
that Maxwell's equations in vacuum
allow for electromagnetic waves, much as the vacuum Einstein equations
support gravitational waves. Indeed, specification of the
$D-1$-form electromagnetic current, $j$, does not yield the
gauge-invariant electromagnetic tensor $F$ because even though 
\be
d*F = j \; ,
\ee
we are free to add a term $F'$ to $F$ satisfying $d*F'=0$. Writing
$F'=dA'$, the term undetermined by the sources satisfies
\be
d*d A' = 0 \; .
\ee
In Lorentz gauge, $d*A' = 0$, we obtain the wave equation,
\be
(*d*d + d*d*)A' = \Delta A' = 0 \; ,
\ee
where $\Delta$ is the d'Alembertian. So we see explicitly that sources determine fields up to
electromagnetic waves. Then, since there are $D-2$ propagating
degrees of freedom, a total of $2(D-2)$ functions on the boundary
must be specified as boundary conditions.

But the understanding that electromagnetic waves are responsible
for the failure of the sources to determine the field also points to a
way out. We know that electromagnetic waves are blocked by a Faraday
cage. Hence if we could place a Faraday cage around the region of interest, all electromagnetic waves coming from the interior would be blocked, and the boundary electromagnetic fields could be attributed to charges and currents living on the cage. Indeed, this is more or less what we will show.
We will see that our proposal amounts to interpreting the
boundary surrounding some region of space as some hypothetical
material --- call it a dual Faraday cage --- which eliminates the
tangential magnetic field. This surface absorbs all incident electromagnetic waves, since
the latter cannot propagate through without a magnetic field component
tangential to the surface. In that sense our approach brings to mind the old Wheeler-Feynman absorber theory, proposed for different reasons, in which electromagnetic waves in the bulk are emitted or absorbed by distant boundary sources \cite{WF1,WF2}.

Our procedure for implementing Machian electromagnetism begins by splitting the Maxwell action, 
\be
S = \int d^Dx\left(-\frac{1}{4g^2}F^2 + J^\alpha A_\alpha\right)\,,
\ee
as integrals over two regions of space, ${\cal M}_{\rm in}$ and ${\cal
  M}_{\rm out}$, separated by a boundary $\Sigma$. In principle $\Sigma$ can be
null, space-like or time-like, although to make contact with common
experience let us take it to be time-like. Call $S_{\rm in}$ and $S_{\rm out}$ the action restricted to an integration over ${\cal M}_{\rm in}$ and ${\cal M}_{\rm out}$ respectively. Now, neither the stationarity of $S_{\rm in}$ nor the stationarity of
$S_{\rm out}$ is sufficient on its own to yield the classical equations of motion because there remains a variation of the field at $\Sigma$, after integration by parts. Following the
action formulation of the membrane paradigm~\cite{membraneaction}, one deals with this left-over variation by adding and subtracting a boundary action on $\Sigma$, 
\be
S=S_{\rm in} + S_\Sigma +S_{\rm out} - S_\Sigma\,,
\ee
with $S_\Sigma$ chosen so that $\delta(S_{\rm in} + S_\Sigma)=0$ classically. Thus, from the point of view of an observer in ${\cal M}_{\rm in}$, $S_\Sigma$ encodes all physical effects of the exterior region. 

Varying the relevant action for this observer yields the bulk equations of motion for the gauge field, $\partial_\alpha F^{\alpha\beta} = -g^2J^\beta$, as well as a boundary term
\be
\delta S = \int_\Sigma d^{D-1}x \left(-\frac{1}{g^2}n_\alpha F^{\alpha\beta}  + \frac{\delta S_\Sigma [A]}{\delta A_{\beta}}\right)\delta A_\beta\,.
\ee
Such a boundary term is usually set to zero by imposing Dirichlet boundary conditions on the variation: $\delta A =0$ on $\Sigma$. However, we can also choose not to fix any definite boundary conditions at $\Sigma$. In the membrane paradigm, one doesn't impose boundary conditions at the horizon because the horizon is regarded as dynamical. Here we don't fix boundary conditions because we want to capture them in terms of matter. Instead, we choose the variation of the boundary action, $S_\Sigma$, to cancel the residual variation from the bulk action:
\be 
\frac{1}{g^2}n_\alpha F^{\alpha\beta}  = \frac{\delta S_\Sigma [A]}{\delta A_{\beta}}\,.
\label{cancel}
\ee
This condition implies that the normal component of the electric field, $E_\bot$, as well as the tangential component of the magnetic field, $B_\|$, are canceled by the boundary action. Thus, as advertised, the boundary acts as a dual Faraday cage preventing incident electromagnetic waves from going through. (We say ``dual" because an ordinary Faraday cage would eliminate $E_\|$ and $B_\bot$; the blocking effect on electromagnetic radiation is identical.)

A general choice for $S_\Sigma$ consistent with all symmetries and with at most two derivatives is
\be
S_\Sigma = \int_\Sigma  d^{D-1}x\left(-\frac{1}{4g^2_b}F^2 + j^\alpha_{(1)} A_\alpha\right)\,,
\ee
where $j_{\rm (1)}$ is a surface current, and where we have allowed for a different coupling constant $g_b$ on the boundary. In this case the matching condition~(\ref{cancel}) gives Maxwell's equations on the boundary,
\be
\partial_\alpha F^{\alpha\beta} = -g^2_b\left(j^\beta_{(1)} - \frac{1}{g^2}n_\alpha F^{\alpha\beta} \right)\,.
\label{condEM1}
\ee
From the point of view of a fiducial observer on $\Sigma$, the last term acts as a current, which we therefore denote by
\be
j^\beta_{(2)} \equiv -\frac{1}{g^2}n_\alpha F^{\alpha\beta} \,.
\label{condEM2}
\ee
In the absence of charge transfer between bulk and boundary,
$j_{(2)}$ is conserved on the boundary, since $\partial_\alpha
j^\alpha_{(2)} \sim n_\alpha J^\alpha = 0$.  It then follows
from~(\ref{condEM1}) that $j_{(1)}$ is separately conserved. 

Specifying the boundary currents $j_{(1)}$ and $j_{(2)}$ completely
determines, through~(\ref{condEM1}) and~(\ref{condEM2}) respectively,
the gauge field on the boundary and its normal derivative, and
therefore encodes the boundary conditions required to obtain a unique
solution for the electromagnetic field in the bulk. As a check on the
number of degrees of freedom in general $D$ dimensions, each current
is a vector in $D-1$ dimensions for a total of $2(D-1)$ components;
but each satisfies a continuity equation which brings the total down
to $2(D-2)$. Sure enough, this is the requisite number of boundary conditions
as a spin-1 massless particle has $D-2$ propagating degrees of
freedom.

Let us sketch the method with a simple
example. Consider a region free of charge
but permeated by a divergence-free electrostatic field, $F^{0i} =
-\partial_i A^0=-\partial_i \phi$, where $\phi$ is the Coulomb
potential. This region is taken to be enclosed by a fictitious time-like
surface $\Sigma$. We would like to determine the required surface
charge densities $j_{(i)}^0\equiv \sigma_{(i)}$, $i=1,2$, on $\Sigma$
that would reproduce the electrostatic field inside. Consulting~(\ref{condEM2}), we find that
the normal component of the electric field is accounted for by $\sigma_{(2)}$:
\be
\sigma_{(2)} = \frac{1}{g^2} \partial_n\phi\,.
\ee
Using the usual equation for the jump in the normal component of $E$ due to
surface charge, $E_\bot^{\rm out}-E_\bot^{\rm in} = g^2 \sigma$, we find incidentally that
$E_\bot^{\rm out}=0$. Similarly, $\sigma_{(1)}$ is determined
by the tangential component through~(\ref{condEM1}):
\be
\sigma_{(1)} = -\frac{1}{g_b^2}\nabla_\|^2\phi -\sigma_{(2)}\,.
\ee
In particular, if the field lines happen to hit $\Sigma$ everywhere normal to the surface, 
then the boundary can be interpreted physically as a thin conductor.

The straightforward simplicity of this derivation should not suggest that it is somehow tautological. For example, had we relied only on the equations of motion --- as opposed to on an action formulation -- we would have been faced with the following problem. We know that surface charges cause a discontinuity in the normal electric field component: $E_\bot^{\rm out} - E_\bot^{\rm in} \sim \sigma$. Now, in order to trade boundary conditions for charges --- $E_\bot^{\rm in}$ for $\sigma$ --- one has to impose some other condition, as otherwise we have only equation relating three variables. But {\it any} independent equation would have done the trick. However, as we shall see, all but one of these conditions fail to possess such desirable properties as the vanishing of global angular momentum, the extinction of incident waves, and consistency with holography and the membrane paradigm. That one condition, which in Machian electromagnetism corresponds to $E_\bot^{\rm out} = 0$, follows naturally, as shown above, from an action principle.  

\subsection{Gravity}

We now turn to Mach's principle in general relativity. The prescription is much the same as for the electromagnetic case studied above. There is, however, one additional subtlety: how do we specify ``where" the matter is if the spacetime geometry has itself not yet been derived? To answer this, let us start from first principles. We begin with a differentiable manifold, ${\cal M}$, a set of points that looks locally like $R^D$. The manifold can be covered by an atlas of charts. Each chart is a set of coordinates, a local map to $R^D$. This then allows us to have a coordinate basis for the tangent space at any given point with basis vectors $\partial / \partial x^a$. But that's all we need: the existence of such a basis is enough for us to be able to specify the components of a rank-two tensor there. 

The bulk $T^{\alpha\beta}$ is such a tensor. There is no need at this stage for an explicit choice of metric, which, being an inner product, is additional structure. (Actually, the stress tensor cannot easily be specified independently of the metric to be endowed on the manifold, because diffeomorphism invariance requires that $T^{\alpha\beta}$ be covariantly conserved. There is therefore a consistency check involved.) Moreover, we will need a boundary, $\Sigma$, whose location can also be specified in terms of its coordinates. On the boundary, we once again use the existence of local coordinates to specify the boundary stress tensor. Put another way, the space-time geometry consists of the {\it triplet} $({\cal M}, g_{\mu \nu}, \Sigma)$. Einstein's version of Mach's principle is then the statement that $g_{\mu \nu}$ can be obtained by specifying bulk and boundary stress tensors on ${\cal M}$ and $\Sigma$, respectively.

Formally, our construction applies equally to space-like and time-like boundaries. However, we want to identify some physical matter living on $\Sigma$, with which a 
bulk observer can potentially interact. The boundary conditions on a time-like surface are more readily interpreted as matter than boundary conditions on a space-like surface. (The null case is also physical but requires separate analysis as the boundary data for the characteristic problem is different~\cite{sachs}. We leave this to future work.) Furthermore a time-like boundary makes contact 
with the Brown-York definition of bulk gravitational stress 
tensor, as well as with the membrane paradigm. As we will see explicitly in 
Section~\ref{examples} the latter is particularly relevant for de Sitter 
space where $\Sigma$ is a surface hovering inside some observer's event horizon. Henceforth, we will assume the boundary to be time-like. Of course to get a unique solution from time-like sources requires that we specify
boundary conditions for all times. This will be the case 
here, just as the membrane description of a black hole horizon is valid for all times for an external observer.

Consider then the Einstein-Hilbert action, including a cosmological term:
\be 
S = \frac{1}{16\pi G_D} \int_{\cal M} d^Dx\sqrt{-g}\left(R-2\Lambda \right)\,,
\label{Sbulk}
\ee
which again splits as integrals over two space-time regions ${\cal M}_{\rm
 in}$ and ${\cal M}_{\rm out}$ separated by $\Sigma$. Once again we add and subtract a boundary action to get $\delta (S_{\rm in} + S_\Sigma+S_{\rm GH})=0$, where the Gibbons-Hawking term,
 \be
 S_{\rm GH} =  \frac{1}{8\pi G_D} \int_\Sigma d^{D-1}x \sqrt{-h}K\,,
 \ee
is necessary to obtain a well-defined variational principle on ${\cal M}_{\rm in}$. (Our convention is that the normal vector to $\Sigma$ is inward-pointing, hence the sign of Gibbons-Hawking.) Thus, from the point of view of an observer in ${\cal M}_{\rm in}$, $S_\Sigma$ encodes the physical effects of the exterior region. Generically we can take $S_\Sigma$ to be a general two-derivative action in $D-1$ dimensions,
\be
S_\Sigma = \frac{1}{16\pi G_{D-1}}\int_{\Sigma} d^{D-1}x \sqrt{-h}\left(^{(D-1)}R-2\lambda\right) + S_\Sigma^{\rm matter}[h]\,,
\label{Sbdy}
\ee
describing intrinsic gravity coupled to a cosmological constant $\lambda$ and boundary matter. This action is a functional of intrinsic boundary quantities only and so leaves the bulk equations unaffected. 

Performing the variation $\delta (S_{\rm in} + S_\Sigma +S_{\rm GH})=0$ then yields the bulk Einstein equations, $G_{\alpha\beta} = -\Lambda g_{\alpha\beta}$, as well as a surface term:
\be
\delta S = \frac{1}{2}\int_\Sigma d^{D-1}x \sqrt{-h}\left\{\frac{1}{8\pi G_D}(K_{\alpha\beta}-Kh_{\alpha\beta})+\frac{2}{\sqrt{-h}}\frac{\delta S_{\Sigma}[h]}{\delta h^{\alpha\beta}} \right\}\delta h^{\alpha\beta}\,.
\label{israel}
\ee 
This is usually set to zero by imposing Dirichlet boundary conditions, $\delta h\vert_\Sigma = 0$. Alternatively, one can choose $S_\Sigma [h]$ to cancel this term. This gives an Israel matching condition, with the only difference being that the extrinsic curvature term for the ``exterior''
region vanishes in this case --- the physics of the exterior region is encoded in the boundary action. 
A similar Israel condition arises in the action description of the membrane paradigm for black holes, this time with the region interior to the black hole
horizon replaced by boundary matter~\cite{membraneparadigm,membraneaction}. The upshot is that~(\ref{israel}) fixes the canonical momentum, $K_{\alpha\beta}-Kh_{\alpha\beta}$, and hence half of the boundary conditions for the gravitational field. 
The vanishing of~(\ref{israel}) is thus the gravitational analogue of~(\ref{cancel}) for electromagnetism --- our boundary acts as a {\it gravitational dual Faraday cage}.
(As in electromagnetism, junction conditions by themselves do not identify the precise relation between boundary condition and boundary sources because an arbitrary choice for the field value beyond the boundary can be made; the action formulation fixes that choice.)

For the choice of $S_{\Sigma}$ given in~(\ref{Sbdy}), the vanishing of the above surface term gives an Einstein equation on the boundary,
\be
^{(D-1)}G_{\alpha\beta} + \lambda h_{\alpha\beta} = \frac{G_{D-1}}{G_D}(Kh_{\alpha\beta}-K_{\alpha\beta}) + 8\pi G_{D-1}T_{\alpha\beta}^{\rm BY-out}\,,
\label{rel0}
\ee
where 
\be
T_{\alpha\beta}^{\rm BY-out}\equiv -\frac{2}{\sqrt{-h}}\frac{\delta S_{\Sigma}^{\rm matter}}{\delta h^{\alpha\beta}}
\ee
is the stress tensor for the boundary matter. This matter stress tensor is recognized as the Brown-York stress tensor for the exterior space-time region ${\cal M}_{\rm out}$, hence the superscript. In analogy with classical mechanics where the energy of a system can be expressed through the Hamilton-Jacobi action as $E=-\partial S/\partial t$, a quasi-local notion of stress-energy for a space-time region can be defined on its boundary as~\cite{brownyork}
\be
T_{\alpha\beta}^{\rm BY} \equiv  \frac{2}{\sqrt{-h}}\frac{\delta S_{\rm class}}{\delta h^{\alpha\beta}} = \frac{1}{8\pi G_D}(K_{\alpha\beta}-Kh_{\alpha\beta}) + \frac{2}{\sqrt{-h}}\frac{\delta S_{\rm reg}}{\delta h^{\alpha\beta}}\,.
\label{BY}
\ee
The regulating term $S_{\rm reg}$ is a local action on the boundary required to cancel potential infrared divergences as the boundary is taken to
infinity or to some horizon, whatever the case may be. For asymptotically locally AdS space-times, such divergences correspond through the AdS/CFT duality to the usual UV divergences in field theory. The cancellation of divergences will end up fixing $G_{D-1}$ and $\lambda$, which at this stage might appear arbitrary. For a 3+1-dimensional bulk, the case of interest, we will find in Section~\ref{ads} the following counter-term action
\be
S_{\rm reg} = \frac{\ell}{16\pi G_4}\int_{\Sigma} d^{3}x \sqrt{-h}\left(^{(3)}R+\frac{4}{\ell^2}\right)\,,
\label{Sreg}
\ee
which is of the form~(\ref{Sbdy}) with $\lambda =-2/\ell^2$ and $G_3=G_4/\ell$, where $\ell$ is the AdS radius~\cite{gubser}.
Substituting this in~(\ref{rel0}) and comparing the result with~(\ref{BY}), 
we indeed see that from the point
of view of an observer in ${\cal M}_{\rm in}$, the effects of the
exterior region can be  encoded with matter on $\Sigma$, whose
corresponding stress tensor $T_{\alpha\beta}^{\rm BY-out}$ is the
Brown-York stress tensor for the gravitational field in ${\cal M}_{\rm
  out}$.
  
The form of~(\ref{rel0}) as Einstein's equations on the boundary suggests that the extrinsic curvature term be interpreted by a boundary observer as a second stress tensor
\be
Kh_{\alpha\beta}-K_{\alpha\beta} =  8\pi G_D T_{\alpha\beta}^{\rm BY-in}\,.
\label{TIR}
\ee
From the discussion above, $T_{\alpha\beta}^{\rm BY-in}$  is just the Brown-York stress tensor for the interior space-time region ${\cal M}_{\rm in}$. 
When there is no matter exchange between boundary and bulk, $T_{\alpha\beta}^{\rm BY-in}$ is conserved on the boundary: 
\be
D_{\alpha}K^\alpha_{\;\beta}-D_\beta K = T^{\rm bulk}_{\alpha\gamma} n^\alpha h^\gamma_{\;\beta} = 0\,.
\ee 
It follows from~(\ref{rel0}) and the Bianchi identity on the boundary, $D^\alpha G_{\alpha\beta} = 0$, that $T_{\alpha\beta}^{\rm BY-out}$ is covariantly conserved as well.

Thus, specifying $T^{\rm BY-in}$ and $T^{\rm BY-out}$ determines $h$
and $K$ through~(\ref{rel0}) and~(\ref{TIR}), and therefore yields a unique
solution for the bulk gravitational field: boundary conditions have
been replaced by boundary sources, as desired. 
Moreover, these stress
tensors have the natural interpretation of describing the stress
energy in the gravitational field for the exterior and interior
space-time regions delimited by $\Sigma$. Thus our proposal realizes
an old suspicion that general relativity could be shown to be Machian by taking into account
not only the matter stress tensor, but also that of the gravitational
field. The natural location for this stress tensor, as Brown and York realized, is on the boundary.

As a check on the counting of degrees of freedom in $D$
dimensions, each stress tensor is a symmetric tensor in $D-1$
dimensions which is covariantly conserved, for a total of $D(D-3) +
2$ degrees of freedom. As argued at the end of Section~\ref{bc}, the Hamiltonian constraint
and the choice of $\Sigma$ give two more conditions, bringing the
total of freely specifiable functions down to $D(D-3)$, in agreement
with the required boundary data for the $D(D-3)/2$ graviton degrees of
freedom.

It is worth commenting that although~(\ref{rel0}) is itself a
differential equation for $h$, in 2+1 dimensions it has a unique
solution up to global identifications --- the Weyl tensor vanishes
identically in 2+1 dimensions, thus curvature is entirely determined
by the Ricci tensor. We therefore have the interesting fact that our version of Mach's principle automatically holds in 2+1 dimensions, even without independent boundary matter. 

In more than four dimensions, equation~(\ref{rel0})
itself requires boundary data, which can be implemented by iterating the
above procedure to lower dimensional surfaces on the boundary. A further complication in higher dimensions, however, is that the counter-term action
in~(\ref{Sreg}) generically includes higher curvature terms. This introduces
higher derivatives in~(\ref{rel0}), which therefore requires more
boundary data. The two-derivative action presented here suffices however for
the 2+1-dimensional boundary of a 3+1-dimensional space-time.

\section{Vanishing Total Angular Momentum} \label{charge}

As argued so far, the addition of boundary matter implements
Einstein's 1918 version of Mach's principle: unique gravitational
field arising from (bulk and boundary) stress energy. With this framework in place, we can ask if other
Machian expectations are also realized. In this brief section we show that
our set-up implies that the total angular momentum of an isolated system is zero. This is in harmony with the Machian precept that motion is only defined in a relative sense.
In fact, the vanishing of all global charges is an immediate consequence of the definition of our boundary term: charges are calculated using the Brown-York stress tensor,
$T_{\alpha\beta}^{\rm BY-in}$, but since we are
adding precisely $-T_{\alpha\beta}^{\rm BY-in}$ on the boundary, the two contributions cancel exactly.

Concretely, corresponding to a given Killing vector $\xi^\alpha$ of
the boundary geometry one can define a conserved current,
\be
j_\alpha = T_{\alpha\beta}^{\rm BY-in}\xi^\beta\,,
\ee
with associated conserved charge:
\be
Q^{\rm bulk} = \oint_B d^{D-2} x \sqrt{\sigma}u^\alpha j_\alpha \,.
\label{chargebulk}
\ee
Here $u^\alpha$ is a time-like vector on $\Sigma$, which is normal to a
$D-2$-dimensional closed surface $B$ with induced metric $\sigma_{\alpha \beta}  =
h_{\alpha \beta}+u_\alpha u_\beta$. From~(\ref{israel}) and~(\ref{TIR}), it is clear that
the addition of $S_{\Sigma}$ gives a second boundary current which
precisely cancels the bulk contribution:
\be
Q^{\rm bulk} + Q^{\rm bdy} = 0\,.
\label{cancelQ}
\ee

For example, consider the angular momentum for the Kerr geometry in 3+1 dimensions. Here we choose $\Sigma$ to be at some large radius where the metric is approximately
\be
ds^2 \approx  -\left(1-\frac{2M}{r}\right)dt^2 -\frac{4J}{r}\sin^2\theta\, dt\,d\phi + \left(1-\frac{2M}{r}\right)^{-1}dr^2 + r^2d\Omega^2\,.
\ee
The ADM angular momentum, corresponding to the Killing vector $\xi^\alpha=(\partial/\partial \phi)^\alpha$, is then
\be
J^{\rm bulk} =  \oint_B d^2x \sqrt{\sigma} u^\alpha T_{\alpha\phi}^{\rm BY-in} = \frac{3J}{8\pi}\int_0^{2\pi} d\phi \int_0^\pi d\theta \sin^3\theta = J\,,
\ee
which is just the statement that the parameter $J$ really is the angular momentum of the space-time. We see that it is indeed given by the Brown-York stress tensor for the interior.
It then follows from~(\ref{cancelQ}) that {\it the boundary rotates in the opposite direction} with
angular momentum $-J$, such that the combined bulk plus boundary sources have precisely zero net angular momentum. Note that a cancellation would not have occurred had one picked a different mapping of the boundary conditions to boundary sources; it is only in our prescription, originating in the action formulation, that conserved charges vanish.

\section{Examples} \label{examples}

In this section we illustrate our construction with several explicit examples. Little has been said thus far about the location of the boundary matter and there is indeed
ample flexibility in this choice. However, space-times with horizons offer natural candidate surfaces: the Rindler horizon of an accelerated observer,
the cosmological horizon in de Sitter space, etc. In these cases the close connection to the membrane paradigm for causal horizons~\cite{causalmembrane} is manifest. For concreteness, we work in 3+1 dimensions throughout this section.

\begin{figure}[ht]
\centering
\includegraphics[width=120mm]{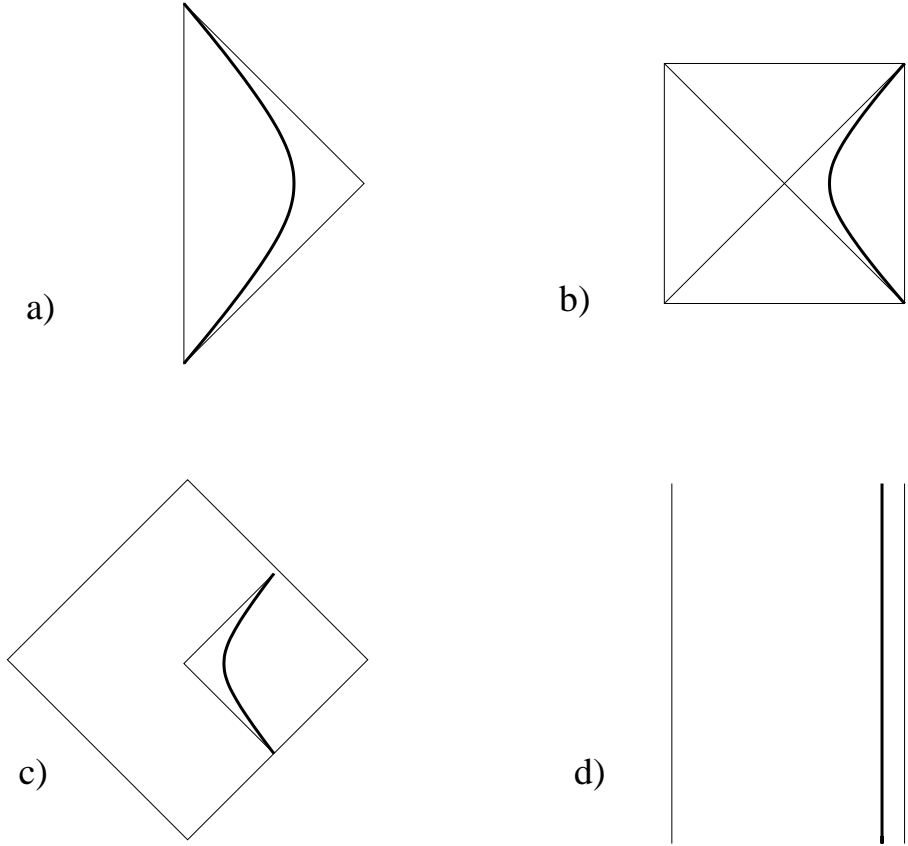}
\caption{Penrose diagrams illustrating our construction for a) Minkowski; b) de Sitter; c) Rindler; d) AdS spaces. The thick curve in each case represents the location
of the boundary stress energy.}
\label{membraneseg}
\end{figure}

\subsection{Minkowski space} \label{mink}

Minkowski space-time has long been considered the quintessential anti-Machian solution to the Einstein field equations. Flat space has no matter, yet there is a well-defined notion of inertia everywhere. Indeed, the same holds true for any of the other vacuum solutions to Einstein's equations. In all these space-times, inertial frames are unambiguously defined and yet there is no (bulk!) matter whatsoever. Thus it would seem that acceleration is defined absolutely and not relatively. Were it not for boundary matter, one could even say that while Newtonian mechanics postulates the existence separately of absolute space and absolute time, Einstein's theory of gravity allows for absolute space-time.

Our proposal draws a different conclusion: it is boundary matter that plays the role of the distant stars. Consider spherically-symmetric coordinates with a fiducial observer sitting at the origin. A natural location for the boundary is the world-volume of a two-sphere at some fixed large radius $r_0$, with topology $R \times S^2$. See Fig.~\ref{membraneseg}a. The boundary geometry is just the Einstein static universe, with the only non-vanishing component of the Einstein tensor given by $G^t_{\;t} = -1/r_0^2$.
Meanwhile, the extrinsic curvature tensor is given by $K^t_{\;t} = 0$;  $K^A_{\;B}  = -\delta^A_{\;B} /r_0$, where $A,B$ denote angular variables on the two-sphere. Substituting in~(\ref{TIR}), we obtain
\bea
\nonumber
(T^{\rm BY-in})^t_{\;t} &=& -\frac{1}{8\pi G_4} \frac{2}{r_0}\;; \\
(T^{\rm BY-in})^A_{\;B} &=& -\frac{\delta^A_{\; B}}{8\pi G_4 r_0} \,.
\eea
The $1/r$ fall-off leads to conserved charges for the space-time which diverge in the limit $r_0\rightarrow\infty$, as seen from~(\ref{chargebulk}). From the Brown-York perspective these can be canceled with appropriate $G_3$ and $\lambda$ to yield a finite $T^{\rm BY-out}$. Indeed, substituting $T^{\rm BY-in}$ and the boundary Einstein tensor in~(\ref{rel0}) gives
\bea
\nonumber
(T^{\rm BY-out})^t_{\;t} &=& \frac{1}{8\pi G_3}\left(-\frac{1}{r_0^2} + \lambda\right) + \frac{1}{8\pi G_4} \frac{2}{r_0}\;; \\
(T^{\rm BY-out})^A_{\;B} &=&\delta^A_{\; B}\left(\frac{\lambda}{8\pi G_3}+ \frac{1}{8\pi G_4 r_0}\right) \,.
\eea
Canceling the diverging terms fixes the cosmological term 
\be
\lambda = -\frac{1}{r_0^2}\,,
\label{cc}
\ee
as well as the intrinsic Newton's constant on the boundary
\be
\frac{G_3}{G_4}= \frac{1}{r_0}\,,
\ee
in agreement with brane-world calculations~\cite{costas}. In particular, we see that gravity decouples in the limit that the boundary is sent to infinity.

Unlike its counterpart in 3+1 dimensions, the Einstein static universe in 2+1 does not require a non-vanishing cosmological term, only dust.
This can be seen directly from the components of the boundary Einstein tensor. Interpreting~(\ref{rel0}) as Einstein's equations on
the boundary, it follows that the cosmological term~(\ref{cc}) and the two boundary stress tensors, $T^{\rm BY-in}$ and $T^{\rm BY-out}$, behave collectively as {\it pressureless dust}, with energy density
\be
\rho_{\rm dust} = \frac{1}{8\pi G_4r_0}\,.
\ee
We may interpret this dust sprinkled on the boundary as determining a
cosmic rest frame for Minkowski space with respect to which
accelerated motion in the bulk is defined. The question ``What does Newton's bucket
spin with respect to in empty space?" finally has a Machian answer: ``The bucket rotates with respect to dust on the surrounding Einstein static universe."

\subsection{de Sitter space} \label{ds}

Another space-time of historical significance for Mach's principle is de Sitter space. It was de Sitter's example of a closed, matter-free cosmological model that eventually led Einstein to abandon Mach's principle. Today de Sitter space is of obvious relevance for early-universe inflation and late-time cosmic acceleration.

Any pair of observers can have two-way communication only within a finite portion of the space-time, the so-called causal diamond. A natural location for our stress energy in this case is the boundary of this region, which is the causal horizon for this observer. (Proponents of the conjectured dS/CFT correspondence have instead focused on ${\cal I}^{\pm}$~\cite{strominger}. While our construction is certainly formally applicable to that case as well, the physical interpretation of matter on a space-like surface is a little unclear.)

Choosing a null boundary leads to some well-known complications: for example, the vector normal to the surface is also on the surface, and the volume element vanishes. Fortunately, the technical subtleties of dealing with a null surface can be circumvented by choosing a time-like surface hovering just inside the horizon, a {\it stretched horizon}, and then taking the limit in which it approaches the true null horizon. See Fig.~\ref{membraneseg}b. This is in precise analogy with the membrane paradigm for black holes~\cite{membraneaction}. (We continue to refer to this time-like surrogate horizon as the stretched horizon even though in de Sitter space, it should probably be called a shrunken horizon.) Unlike the black hole membrane paradigm, for which only those observers who remain outside the black hole see a membrane, here every observer has its own stretched horizon. The physical interpretation of the membrane in such a scenario probably involves some form of observer complementarity \cite{obscomp}.

The metric in the causal diamond is given by
\be
ds^2= - (1-\Lambda r^2)dt^2 + \frac{dr^2}{1-\Lambda r^2} + r^2d\Omega^2\,,
\label{desitter}
\ee
with our fiducial observer sitting at  $r=0$, and his causal horizon located at $r=\Lambda^{-1/2}$. 
The stretched horizon, on which boundary matter will be pasted, is defined as the time-like surface $r=r_0$ such that the lapse function
satisfies
\be
\alpha\equiv \sqrt{1-\Lambda r_0^2} \ll 1\,. 
\ee
At the end of the day we will be interested in the limit $\alpha\rightarrow 0$, in which the stretched horizon merges with the true horizon.

Substituting the extrinsic curvature components $K^t_{\;t} = \Lambda r_0/\alpha$ and $K^A_{\; B} = -\delta^A_{\;B}\alpha/r_0$ in~(\ref{TIR})
yields the following ``in" stress tensor
\bea
\nonumber
(T^{\rm BY-in})^t_{\;t} &=& -\frac{1}{8\pi G_4} \frac{2\alpha}{r_0} \equiv -\frac{\theta}{8\pi G_4}\;; \\
(T^{\rm BY-in})^A_{\;B} &=& \frac{\delta^A_{\; B}}{8\pi G_4} \left[\frac{\alpha}{r_0} + \frac{1}{\alpha r_0}\right] \equiv  \frac{\delta^A_{\; B}}{8\pi G_4} \left[\frac{\theta}{2} + g\right]\,,
\eea
where $\theta$ is the expansion parameter for a congruence of radial null geodesics, while $g \equiv r_0^{-1}\alpha^{-1}$ is the proper surface gravity. Thus, exactly as in the membrane paradigm for black holes, we recognize the stress tensor of a Newtonian fluid with energy density $\rho = \theta/8\pi G_4$, pressure $P = g/8\pi G_4$, and bulk viscosity $\zeta = -1/16\pi G_4$. Moreover, this fluid satisfies a host of non-relativistic equations, such as the Navier-Stokes equation, Ohm's law, and Joule's law~\cite{membraneparadigm}. We see therefore that de Sitter horizons also possess a membrane interpretation.

The $1/\alpha$ divergence in the pressure can be canceled by counterterms with suitable choices of $G_3$ and $\lambda$ in~(\ref{rel0}) to yield a finite $T^{\rm BY-out}$. Substituting in~(\ref{rel0}) the above $T^{\rm BY-in}$, as well as the boundary Einstein tensor with components $G^t_{\; t} = -1/r_0^2$ and $G^A_{\; B} = 0$, we find
\bea
\nonumber
(T^{\rm BY-out})^t_{\;t} &=& \frac{1}{8\pi G_3}\left(-\frac{1}{r_0^2} + \lambda\right) + \frac{1}{8\pi G_4} \frac{2\alpha}{r_0}\;; \\
(T^{\rm BY-out})^A_{\;B} &=&\delta^A_{\; B}\left[\frac{\lambda}{8\pi G_3}- \frac{1}{8\pi G_4}\left(\frac{\alpha}{r_0}+\frac{1}{\alpha r_0}\right)\right] \,.
\eea
It is easily seen that the required cosmological term and gravitational constant are
\be
\lambda = \frac{1}{r_0^2}\;;\qquad \frac{G_3}{G_4} = \frac{\alpha}{r_0}\,.
\ee
Gravity therefore decouples in the limit that the stretched and true horizons merge, analogous to our Minkowski analysis. 

As in Minkowski space, the boundary geometry is once again Einstein's static universe in 2+1 dimensions. All sources on the boundary, including the cosmological term, therefore add up to an effective dust fluid with surface energy density
\be
\rho_{\rm dust} = \frac{1}{8\pi G_4r_0\alpha}\,.
\ee
This diverges as $\alpha\rightarrow 0$ since the dust fluid has infinite proper acceleration in this limit. The divergence is familiar from the black hole membrane paradigm for which the energy density on the membrane also diverges in the limit that the time-like stretched horizon approaches the null event horizon.

\subsection{Rindler space}

Our next example is Rindler space, describing a uniformly accelerated observer in Minkowski space. The Rindler trajectory starts at ${\cal I}^-$ and ends at ${\cal I}^+$, thereby defining a causal horizon. A natural location for our stress tensor is the stretched horizon hovering over this Rindler horizon, as shown in Fig.~\ref{membraneseg}c. 

The construction is very similar to the previous example since the near-horizon geometry of de Sitter space is Rindler space. To see this explicitly, we introduce a new radial coordinate $z \equiv \alpha/\Lambda r$ and a dimensionless time $\tilde{t}\equiv \sqrt{\Lambda}t$, in terms of which the line element~(\ref{desitter}) takes the form
\be
ds^2 = (1+\Lambda z^2)^{-1}\left\{-z^2d\tilde{t}^2+dz^2+\frac{1}{\Lambda}d\Omega^2\right\}\,.
\ee
In the limit $\Lambda\rightarrow 0$, the metric on the sphere becomes approximately that of a two-dimensional plane, and we have
\be
ds^2 = -z^2d\tilde{t}^2 + dz^2 + dx_idx^i\,,
\ee
which describes Rindler space. In this coordinate system, the trajectory of the Rindler observer is $z={\rm constant}$, the causal horizon is at $z=0$, and
we will denote by $z_0\ll 1$ the location of the stretched horizon. 

In terms of the original coordinates the above limit corresponds to taking $\Lambda\rightarrow 0$, $\alpha\rightarrow 0$, such that $\alpha/\sqrt{\Lambda}$ is finite. Thus we can easily obtain all the desired quantities by taking this limit of our results for de Sitter space. For instance we deduce that the ``in" stress tensor has components $(T^{\rm BY-in})^t_{\;t} = 0$ and $(T^{\rm BY-in})^i_{\;j} = \delta^i_{\;j} /8\pi G_4z_0$. Moreover, $\lambda\rightarrow 0$ and $G_3\rightarrow 0$, but such that $\lambda/8\pi G_3 \rightarrow 1/8\pi G_4 z_0$. Once again one can think of the total stress energy plus cosmological term as a boundary dust fluid with energy density
\be
\rho_{\rm dust} = \frac{1}{8\pi G_4z_0}\,,
\ee
which, as before, diverges in the limit $z_0\rightarrow 0$ where the stretched horizon merges with the true horizon.

\subsection{Anti-de Sitter space} \label{ads}

Finally, we consider four-dimensional anti-de Sitter space. In global coordinates, the AdS$_4$ line element can be written as
\be
ds^2 = -\left(1+\frac{r^2}{\ell^2}\right)dt^2 + \left(1+\frac{r^2}{\ell^2}\right)^{-1}dr^2 + r^2d\Omega^2\,,
\ee
with $\ell$ the AdS radius. A natural location for our boundary stress energy in this case is a large sphere at $r=r_0\gg \ell$. See Fig.~\ref{membraneseg}d. 
From~(\ref{TIR}) the components of the ``in" stress tensor are straightforwardly calculated:
\bea
\nonumber
(T^{\rm BY-in})^t_{\;t} &=& -\frac{1}{4\pi G_4\ell} \sqrt{1+\ell^2/r_0^2} \approx -\frac{1}{4\pi G_4\ell} -\frac{\ell}{8\pi G_4 r_0^2} \;; \\
(T^{\rm BY-in})^A_{\;B} &=& -\frac{\delta^A_{\;B}}{8\pi G_4\ell}\frac{2+\ell^2/r_0^2}{\sqrt{1+\ell^2/r_0^2}} \approx   -\frac{\delta^A_{\;B}}{4\pi G_4\ell} \,.
\eea
The constant and $r_0^{-2}$ terms both result in infrared divergences in global charges such as the AdS energy and must be regulated with appropriate counterterms. Since the boundary is once again $R \times S^2$, the intrinsic Einstein equations~(\ref{rel0}) yield
\bea
\nonumber
(T^{\rm BY-out})^t_{\;t} &=& \frac{1}{8\pi G_3}\left(-\frac{1}{r_0^2} + \lambda\right) + \frac{1}{4\pi G_4\ell} +\frac{\ell}{8\pi G_4 r_0^2}\;; \\
(T^{\rm BY-out})^A_{\;B} &=&\delta^A_{\; B}\left(\frac{\lambda}{8\pi G_3} + \frac{1}{4\pi G_4\ell}\right) \,.
\eea
The requirement that the divergent terms drop out uniquely fixes the cosmological term and gravitational coupling constant on the boundary:
\be
\lambda = -\frac{2}{\ell^2}\;;\qquad G_3 = \frac{G_4}{\ell}\,,
\ee
which confirms~(\ref{Sreg}) and is in perfect agreement with earlier calculations in AdS/CFT~\cite{gubser,vj,costas}. 

We regard the agreement with AdS/CFT as an important validation of our approach. And yet, at first sight, Mach's principle seems to fly against one of the great achievements of string theory: the unification of particle physics with gravity. Indeed our proposal 
treats gravity and matter fields on different footings since, in the final analysis, gravity, unlike matter, is determined also by sources at the boundary. Why then do our results agree, in AdS, with those of AdS/CFT? The answer might be that Mach's principle may already be embodied in string theory, where it would appear in the guise of open/closed string duality, the idea that the world-sheets of tree-level closed string diagrams are conformally equivalent to planar open string loops. From the world-sheet point of view, we are replacing the far infrared contribution to closed string propagation with loop diagrams in the dual open string channel~\cite{openclosed}. This interpretation points to a generalization of Mach's principle in which not just gravity but all fields corresponding to closed string excitations have matter counterparts at the boundary. While in string theory such a holographic duality has been realized only in certain backgrounds, in particular in AdS, our construction provides evidence that it may be more generally true.

In the context of string theory, it is worth mentioning an intriguing manifestation of Mach's principle in Ho\v rava's Cherns-Simons proposal for M-theory~\cite{horava}. Here the emergence of macroscopic space-time requires turning on a large number of Wilson lines in the gauge theory. The inertia of a propagating excitation is then understood as arising from interactions with this background ``matter."

\section{Comparison with Other Machian Proposals} \label{comp}

There have been numerous previous attempts to reconcile Machian ideas with a relativistic theory of gravity. 
For the purpose of contrasting our framework with existing ideas we will focus our attention on
a few key proposals. In terms of strategy, these have either
i) imposed a selection rule to remove unwanted solutions to the Einstein equations, based on some criterion of Machianity; or
ii) sought an alternative to Einstein's theory with the hope of fulfilling some Machian expectations. 

The classic example of a ``selection rule," proposed by Einstein himself and later pursued by Wheeler~\cite{wheelerclosed}, is the requirement that the universe have closed spatial topology. This condition removes the
need for spatial boundary or asymptotic conditions on the three-metric. Moreover, there is no net global charge in a closed universe, thereby fulfilling another Machian expectation. The prime example of such a universe is Einstein's static universe with global topology $R\times S^3$. 

The Einstein-Wheeler universe has been the subject of ample literature, from which we draw three main objections.
First, in a technical sense the need for boundary data is not entirely obviated since a unique solution to the Einstein equations still
requires some extra data specified, {\it e.g.}, on an initial surface. Intuitively this is because, even in a spatially closed universe, one can still add a generic superposition of standing gravity waves each satisfying the appropriate closed boundary conditions; the effect of compactifying spatial dimensions is to discretize the wave numbers of gravity waves, not to eliminate them altogether. Second, there is potentially a problem with causality. Since the radius of curvature of our universe 
is much larger than the horizon size, as indicated by cosmological observations, how can the local notion of inertia be determined by conditions beyond the observable universe? A third objection, which inevitably afflicts any ``selection rule" approach, is that it is nothing more than an {\it ad hoc} patch. For instance, the Einstein-Wheeler prescription forbids Minkowski space since its spatial slices are non-compact, to the relief of Machian proponents, while permitting $R\times T^3$ with {\it arbitrarily large} torus. 

The second approach gives up on the idea that Machianity is somehow realized within the framework of general relativity and proposes instead some
modified theory of gravity. The poster-child example in this category, and the theory perhaps best motivated by subsequent developments in particle physics, is the scalar-tensor theory of Brans and Dicke~\cite{bransdicke}. At its core is the principle that a Machian universe must satisfy
 \be
 \frac{G_4M}{R}\sim 1\,,
\label{BD}
 \ee
where, heuristically, $M$ and $R$ are the mass and radius of the observable universe. This relation is deduced from inertial induction arguments~\cite{sciamainert,dickeinert}  --- essentially the requirement that accelerating an observer with respect to the distant stars be physically equivalent to accelerating all of the other matter in the universe. For instance, in the Lense-Thirring effect~\cite{LT,brillcohen}, inertial frames inside a rotating shell of mass $M$ and radius $R$ are dragged at the same angular velocity as the shell in the limit where~(\ref{BD}) is satisfied.

One can view~(\ref{BD}) as a constraint on the matter of the universe, requiring special initial or boundary conditions. From a modern perspective,~(\ref{BD}) follows from the near flatness of our universe, which in turn traces back to early-universe inflation, {\it i.e.}, to initial conditions. The key insight of Brans and Dicke is that such a relation can be dynamically satisfied if Newton's constant $G_4$ is time-dependent, continuously adjusting its value according to the matter content of the universe. But while~(\ref{BD}) may indeed be an essential property of a Machian universe, at the end of the day Brans-Dicke theory does not escape the need for boundary/initial data. In Einstein frame, the Brans-Dicke scalar field governing the space-time evolution of $G_4$ is merely an extra matter field coupled to gravity --- and gravity still requires boundary data. Indeed, any gravitational theory that asks for solutions to a differential equation cannot avoid the problem of boundary conditions.

At a spiritual level our proposal is perhaps closest to the boundary matter contemplated by Einstein for some time in 1916~\cite{ds1916,einstein1917}. The idea is to suppose that the metric at large distances from the matter distribution assumes the degenerate form
\be
g_{\mu 0} \rightarrow \infty \;;\qquad g_{ij} \rightarrow 0\,.
\ee
This form of the metric, being invariant under coordinate transformations to arbitrarily accelerated coordinate frames (keeping $t$ fixed), ensures the asymptotic equivalence of all observers~\cite{peeblesbook}. Einstein reasoned that the matter distribution responsible for the transition from a nearly flat metric in our neighborhood to the above degenerate form cannot be the distant stars, for otherwise the large change in the gravitational field would lead to unacceptably large redshift of their spectra. Instead it must be attributed to unseen boundary matter beyond the matter distribution. However, the existence of such dark matter seemed unappealing at the time, and Einstein dropped this idea in favor of his closed cosmological model.

\section{Summary and Outlook} \label{summary}

In this paper we have argued that there exists an interpretation of Mach's principle --- unique gravitational field arising from a given matter distribution --- that can be made to hold in general relativity. Inertial frames are determined by the metric. The metric
is determined by the bulk stress tensor plus boundary
conditions. We have shown that the latter can be replaced by appropriate boundary stress energy
on a hypersurface. Thus a unique metric is obtained provided one specifies both
bulk and boundary stress energy. In concrete terms this boundary matter plays the role
of the distant stars for bulk observers, and heuristically selects a frame with respect to which inertial and accelerated motion have meaning. Our specific proposal is validated by consistency with several other prominent themes in gravity ranging from the membrane paradigm to the idea of a boundary stress tensor for gravitational energy to holography. It satisfies many of the properties that have previously been considered under the rubric of Mach's principle --- distant matter as the source of inertial frames, the absence of global angular momentum, and the relativity of all motion --- and it does so in the most conservative way, without harming either Einstein's equations or their solutions.

By realizing Mach's principle using boundary matter, we see paradoxically that the metric is both dispensable and essential. Much as the Wheeler-Feynman absorber theory did away with electromagnetic fields in favor of charges and currents, so in a Machian reformulation we can understand Einstein's theory without recourse to a metric. But the resultant non-local acausal action-at-a-distance theory is so unwieldy and impractical, that we appreciate exactly why a local metric is such a good thing to have. This is a familiar story in holography: locality of the bulk is hard to see from the perspective of the boundary theory, but manifest in the bulk.

Our framework for implementing Mach's principle in general relativity opens many unexplored avenues for further research.

\begin{itemize}

\item {\it Straightening out arbitrary worldlines} \\
The underlying idea of our proposal is that non-inertial motion can be defined with respect to the boundary. It would be very interesting to test the relativity of acceleration by investigating whether an arbitrary curved trajectory in the bulk can be made inertial --- the worldline straightened out into a geodesic --- by suitably adjusting the boundary matter. This would achieve complete relativity of motion, here the motion of a bulk observer with respect to boundary matter.

\item {\it Total relativity} \\
Specifying bulk and boundary stress energy singles out a metric and, through it, a set of preferred worldlines, namely the inertial ones. This is, loosely speaking, reminiscent of spontaneous symmetry breaking in field theory and hints at a larger symmetry group underlying a complete
relativitization of motion \cite{totalrelativity}. Our framework suggests that the boundary is the natural location for hidden symmetries.
There are indeed examples in general relativity of transformations that leave the asymptotic geometry invariant; it is tempting to speculate that the so-called Bondi-Metzner-Sachs and Spi groups for asymptotically flat spaces might be relevant here.

\item{\it Mach's principle in 2+1 dimensions} \\
Something intriguing happens in three dimensions: Mach's principle automatically holds.
In three dimensions, there are no gravitational waves, the Weyl tensor vanishes identically, and the Riemann tensor is completely determined by the Ricci tensor. Moreover,
there is no freedom to choose gauge-invariant physical boundary
conditions with which to solve the Einstein equations.
Hence Einstein's 1918 version of Mach's principle already holds in
2+1 dimensional gravity. Nevertheless, there exist different types of geometries 
with the same bulk stress tensor. For example, besides AdS$_3$ there are also BTZ black holes
of various masses and angular momenta, all
obtainable via identifications on the universal covering space of AdS. 
It may be worthwhile to study these in the context of our realization of Mach's principle.

\item {\it Observational tests} \\
We described in detail the application of our formalism to de Sitter space. Given the mounting evidence for a small cosmological constant, it is likely that our universe will asymptote to de Sitter space. It is therefore imperative to study potential observable
consequences of the boundary-matter description; after all, the black hole membrane paradigm was designed to perfectly mimic the observations of a specific kind of observer.

\end{itemize}

\noindent
To conclude, we are hopeful that, just as the membrane paradigm was valuable in the study of black holes, so our boundary approach to Mach's principle may prove fruitful for better understanding some of the foundations of general relativity.

\bigskip
\noindent
{\bf Acknowledgments}

\noindent
We are grateful to the Chinatown bus service between Boston and New York City for having made this collaboration possible. Our gratitude goes also to B.~Greene, P.~Ho\v rava, P.J.E.~Peebles, L.~Smolin, A.~Starinets, E.~Verlinde, and F.~Wilczek for many insightful discussions. The research of J.K. at Perimeter Institute is supported in part by the Government of Canada through NSERC and by the Province of Ontario through MEDT. The research of M.P. at Columbia University was supported in part by a Frontiers of Science fellowship and partially by DOE grant DE-FG02-92ER40699. The authors also acknowledge support of Foundational Questions Institute (FQXi) grant RFP1-06-20.

\end{document}